\documentclass[11pt]{article}
\usepackage{amsthm}
\usepackage{amssymb}
\usepackage{amsmath}
\usepackage{mathptmx}

\newcommand{\vnote}[1]{}

\usepackage{fullpage}
\usepackage[utf8]{inputenc}
\usepackage{mathtools}
\DeclarePairedDelimiter{\floor}{\lfloor}{\rfloor}
\DeclarePairedDelimiter{\ceil}{\lceil}{\rceil}
\usepackage[pagebackref]{hyperref}
\theoremstyle{plain}

\renewcommand{\le}{\leqslant}
\renewcommand{\ge}{\geqslant}

\newcommand{\mixstr}{mixed}

\newtheorem{theorem}{Theorem}
\newtheorem{Lemma}[theorem]{Lemma}

\theoremstyle{definition}
\newtheorem{df}{Definition}

\theoremstyle{remark}

\newcommand{\F}{\mathbb{F}}

\newcommand{\Fail}{\operatorname{Fail}}
\newcommand{\rep}{\operatorname{rep}}
\newcommand{\RS}{\mathsf{RS}}
\title{Efficient Low-Redundancy Codes for \\ Correcting Multiple Deletions\thanks{This paper was presented at the 2016 ACM-SIAM Symposium on Discrete Algorithms~\cite{self}. A journal version appeared in IEEE Transactions on Information Theory~\cite{self2}.}}
\author{Joshua Brakensiek\thanks{Department of Mathematical Sciences, Carnegie Mellon University, Pittsburgh, PA 15213, USA. Email: {\tt jbrakens@andrew.cmu.edu}. Research supported in part by an REU supplement to NSF CCF-0963975.} \and Venkatesan Guruswami\thanks{Computer Science Department, Carnegie Mellon University, Pittsburgh, PA 15213. Email: {\tt guruswami@cmu.edu}. Research supported in part by NSF grants CCF-0963975 and CCF-1422045.} \and Samuel Zbarsky\thanks{Department of Mathematical Sciences, Carnegie Mellon University, Pittsburgh, PA 15213, USA. Email: {\tt szbarsky@andrew.cmu.edu}}}
\date{}

\begin{document}

\maketitle
\thispagestyle{empty}

\begin{abstract}

We consider the problem of constructing binary codes to recover from
$k$--bit deletions with efficient encoding/decoding, for a fixed
$k$. The single deletion case is well understood, with the
Varshamov-Tenengolts-Levenshtein code from 1965 giving an asymptotically optimal
construction with $\approx 2^n/n$ codewords of length $n$, i.e., at
most $\log n$ bits of redundancy. However, even for the case of two
deletions, there was no known explicit construction with redundancy
less than $n^{\Omega(1)}$.

For any fixed $k$, we construct a binary code with $c_k \log n$
redundancy that can be decoded from $k$ deletions in $O_k(n \log^4 n)$
time. The coefficient $c_k$ can be taken to be $O(k^2 \log k)$, which
is only quadratically worse than the optimal, non-constructive bound
of $O(k)$. We also indicate how to modify this code to allow for a
combination of up to $k$ insertions and deletions.

\end{abstract}

\section{Introduction}
A $k$-bit binary deletion code of length $N$ is some set of strings
$C\subseteq\{0,1\}^N$ so that for any $c_1,c_2\in C$, the longest
common subsequence of $c_1$ and $c_2$ has length less than $N-k$. For
such a code, a codeword of $C$ can be uniquely identified from any of
its subsequences of length $N-k$, and therefore such a code enables recovery from $k$ adversarial/worst-case deletions.

In this work, we are interested in the regime when $k$ is a fixed constant, and the block length $N$ grows. Denoting by $\mathsf{del}(N,k)$ the size of the largest $k$-bit binary deletion code of length $N$, it is known that 
\begin{equation}
\label{eq:del-combin}
 a_k \frac{2^N}{N^{2k}} \le \mathsf{del}(N,k) \le A_k \frac{2^N}{N^k} 
\end{equation}
for some constants $a_k > 0$ and $A_k < \infty$ depending only
$k$~\cite{Levenshtein66}.  New upper bounds on code
size for a fixed number of deletions that improve over
\cite{Levenshtein66} were recently obtained in \cite{CullinaK14}.

For the special case of $k=1$, it is known that $\mathsf{del}(N,1) = \Theta(2^N/N)$. The Varshamov-Tenengolts code~\cite{{VarshamovTenengolts65}} defined by 
\begin{equation}
\label{eq:VT-code}
\Big\{(x_1,\ldots,x_N)\in\{0,1\}^N \ \mid \ \sum_{i=1}^N ix_i\equiv 0\pmod{(N+1)}\Big\}
\end{equation}
is known to have size at least $2^N/(N+1)$, and Levenshtein~\cite{Levenshtein66} shows that this code is capable of correcting a single deletion. An easy to read exposition of the deletion correcting property of the VT code can be found in the detailed survey on single-deletion-correcting codes~\cite{Sloane02}.

The bound \eqref{eq:del-combin} shows that the asymptotic number of
redundant bits needed for correcting $k$-bit deletions in an $N$-bit
codeword is $\Theta(k \log N)$ (i.e., one can encode $n = N - \Theta(k
\log N)$ message bits into length $N$ codewords in a manner resilient
to $k$ deletions). Note that the codes underlying this result are
obtained by an exponential time greedy search --- an efficient
construction of $k$-bit binary deletion codes with redundancy
approaching $O(k \log N)$ was not known,
except in the single-deletion case where the VT code gives a solution
with optimal $\log N + O(1)$ redundancy.

The simplest code to correct $k$ worst-case deletions is the $(k+1)$-fold
repetition code, which simply repeats each bit $(k+1)$ times, mapping $n$ message bits to $N = (k+1)n$ codewords
bits. It thus has $\frac{k}{k+1} N$ redundant bits. A generalization
of the VT code for the case of multiple deletions was proposed in
\cite{HF02} and later proved to work in \cite{Ghaffaretal12}. These
codes replace the weight $i$ given to the $i$'th codeword bit in the
check constraint of the VT code \eqref{eq:VT-code} by a much larger
weight, which even for the $k=2$ case is related to the Fibonacci
sequence and thus grows exponentially in $i$. Therefore, the
redundancy of these codes is $\Omega(N)$ even for two deletions, and
equals $c_k N$ where the constant $c_k \to 1$ as $k$ increases. Some
improvements were made for small $k$ in \cite{Paluncicetal12}, which
studied run-length limited codes for correcting insertions/deletions,
but the redundancy remained $\Omega(N)$ even for two deletions.

Allowing for $\Theta(N)$ redundancy, one can in fact efficiently correct a
constant \emph{fraction} of deletions, as was shown by Schulman and
Zuckerman~ \cite{SchulmanZuckerman99}. This construction was improved
and optimized recently in \cite{GW-ieeetit}, where it was shown that
one could correct a fraction $\zeta > 0$ of deletions with
$O(\sqrt{\zeta} N)$ redundant bits in the encoding. One can deduce
codes to correct a constant $k$ number of deletions with redundancy
$O_k(\sqrt{N})$\footnote{We use the notation $O_k$ to indicate that the constant may depend on $k$.} using the methods of \cite{GW-ieeetit} (we will hint
at this in Section~\ref{sec:overview}). 

In summary, despite being such a natural and basic problem, there were
no known explicit codes with redundancy better than $\sqrt{N}$ even
to correct from two deletions. Our main result, stated formally as
Theorem~\ref{thm:main} below, gives an explicit construction with
redundancy $O_k(\log N)$ for any fixed number $k$ of deletions, along with a near-linear time decoding algorithm. 

For simplicity, the above discussion focused on the problem of
recovering from deletions alone. One might want codes to recover from
a combination of deletions and insertions (i.e., errors under the edit
distance metric). Levenshtein~\cite{Levenshtein66} showed that any
code capable of correcting $k$ deletions is in fact also capable of
correcting from any combination of a total of $k$ insertions and
deletions.  But this only concerns the combinatorial property
underlying correction from insertions/deletions, and does not
automatically yield an algorithm to recover from insertions/deletions
based on a deletion-correcting algorithm. For our main result, we are
able to extend our construction to efficiently recover from an arbitrary
combination of worst-case insertions/deletions as long as their total number is at most $k$.

\subsection{Our result}
\label{sec:intro-result}

In this work,  we construct, for each fixed $k$, a binary code
of block length $N$ for correcting $k$ insertions/deletions on which all relevant operations can be done in
polynomial (in fact, near-linear) time and that has $O(k^2 \log k \log
N)$ redundancy. We stress that this is the first efficient construction with
redundancy smaller than $N^{\Theta(1)}$ even for the $2$-bit deletion case. For
simplicity of exposition, we go through the details on how to construct an
efficient deletion code, and then indicate how to modify it to turn it
into an efficient deletion/insertion code. 

\newcommand{\enc}{\mathsf{Enc}}
\newcommand{\dec}{\mathsf{Dec}}
\begin{theorem}
\label{thm:main}
Fix an integer $k \ge 2$, For all sufficiently large $n$, there exists a code length $N \le \allowbreak n + \allowbreak O(k^2 \log k \log n)$, an injective encoding map $\enc : \{0,1\}^n \to \{0,1\}^N$ and a decoding map $\dec : \{0,1\}^{N-k} \to \{0, 1\}^n \cup \{\Fail\}$ both computable in $O_k(n(\log n)^4)$ time, such that for all $s \in \{0, 1\}^n$ and every subsequence $s' \in \{0,1\}^{N-k}$ obtained from $\enc(s)$ by deleting $k$ bits, $\dec(s') = s$. 
\end{theorem}

Note that the decoding complexity in the above result has a FPT
(fixed-parameter tractable) type dependence on $k$, and a near-linear
dependence on $n$.

Our encoding function in Theorem~\ref{thm:main} is non-linear. This is
inherent; even to correct a single deletion, linear codes must have
rate at most $1/2$~\cite{Ghaffaretal07}. The published versions of
this paper~\cite{self,self2} claimed in an appendix that linear
$k$-bit deletion codes must have rate at most $\approx 1/(k+1)$ (which
is achieved by the trivial $(k+1)$-fold repetition code); this claim
is, however, false as pointed out to us by Kuan Cheng and Xin Li.

\subsection{Our approach}
\label{sec:overview}
We describe at a high level the ideas behind our construction of $k$-bit binary deletion codes with logarithmic redundancy. The difficulty with the deletion channel is that we don't know the location of deletions, and thus we lose knowledge of which position a bit corresponds to. Towards identifying positions of some bits in spite of the deletions, we can break a codeword into blocks $a_1,a_2,\dots,a_m$ of length $b$ bits each, and separate them by introducing dummy buffers (consisting of a long enough run of $0$'s, say). If only $k$ bits are deleted, by looking for these buffers in the received subsequence, we can identify all but $O(k)$ of the blocks correctly (there are some details one must get right to achieve this, but these are not difficult). If the blocks are protected against $O(k)$ errors, then we can recover the codeword. This can be achieved by a ``syndrome hash" of $O(kb)$ bits knowledge of which enables correction of those $O(k)$ block errors.
In terms of redundancy, one needs at least $m$ bits for the buffers, and at least $\Omega(kb) \ge b$ bits to correct the errors in the blocks.  As $mb =n$, such a scheme needs at least $\Omega(\sqrt{n})$ redundant bits. Using this approach, one can in fact achieve $\approx \sqrt{kn}$ redundancy; this is implicit in \cite{GW-ieeetit}.

To get lower redundancy, our approach departs from the introduction of explicit buffers, as they use up too many redundant bits. Our key idea is to use patterns that occur frequently in the string {\em themselves} as implicit buffers, so we have no redundancy wasted for introducing buffers. For example, if the substring ``00110111'' occurs frequently in the string, we can use it as a buffer to separate the string into short blocks. Since an adversary could foil our approach by deleting a bit that is part of an implicit buffer, we use multiple implicit patterns and form a separate ``hash" for each pattern (the hash will protect the intervening blocks against $k$ errors). Since some strings have very few suitable short patterns (such as the all $0$'s string), we first use a {\em pattern enriching} encoding procedure to ensure that there are sufficiently many patterns. The number of implicit patterns is enough so that less than half of them are corrupted by an adversary for any choice of $k$ deletions. Then, we can decode the string using each pattern and take the majority vote of the resulting decodings. The final codeword bundles the {\em pattern rich} string, a hash describing the pattern enriching procedure (allowing one to recover the original string from the pattern rich string), and the hash for each pattern. 

The two hashes are protected with a less efficient $k$-bit deletion code (with $o(n)$ redundancy) and the decoding procedure begins by recovering them correctly.  Consider a pattern $p$ none of whose occurrences in the pattern rich part of the codeword are affected by the $k$-bit deletion pattern. Given the correct value of the hash associated with this pattern $p$, one can correct the at most $k$ errors in the intervening blocks that are demarcated by occurrences of $p$. 
The algorithm attempts such a recovery procedure for every choice of $p$ (of certain length), and outputs the string $s$ that occurs as the result in a majority of such decodings; the existence of such a majority string is guaranteed by the fact that more than half the patterns tried do not incur any of the $k$ deletions. This implies that $s$ must equal the correct pattern rich portion of the codeword. Finally, the original message is recovered by inverting the pattern enriching procedure on $s$ using knowledge of the corresponding portion of the hash.

\subsection{Deletion codes and synchronization protocols}
\label{sec:sync}
A related problem to correcting under edit distance is the problem of
synchronizing two strings that are nearby in edit distance or {\em
  document exchange}~\cite{CPSV00}. The model here is that Alice holds
a string $x \in \{0,1\}^n$ and Bob holds an arbitrary string $y$ at
edit distance at most $k$ from $x$ --- for simplicity let us consider
the deletions only case so that $y \in \{0,1\}^{n-k}$ is a subsequence
of $x$. The existential result for deletion codes implies that there
is a short message $g(x) \in \{0,1\}^{O(k \log n)}$ that Alice can
send to Bob, which together with $y$ enables him to recover
$x$ (this is also a special case of a more general communication problem considered in \cite{Orlitsky93}). However, the function $g$ takes exponential time to compute. We
note that if we had an efficient algorithm to compute $g$ with output length
$O(k \log n)$, then one can also get deletion codes with small
redundancy by protecting $g(x)$ with a deletion code (that is shorter
and therefore easier to construct).  Indeed, this is in effect what
our approach outline above does, but only when $x$ is a pattern rich
string. Our methods don't yield a deterministic protocol for this
problem when $x$ is arbitrary, and constructing such a protocol with
$n^{o(1)}$ communication was open until Belazzougui \cite{B15} constructed a deterministic protocol with $O(k^2 + k\log^2(n))$ redundancy. See the next section for more details. 

If we allow randomization, sending a random hash value $h(x)$ of $O(k
\log n)$ bits will allow Bob to correctly identify $x$ among all
possible supersequences of $y$; however, this will take $n^{O(k)}$
time. Randomized protocols that enable Bob to efficiently recover $x$
in near-linear time are known, but these require larger hashes of size
$O(k \log^2 n \log^{\ast} n)$~\cite{Jowhari12} or $O(k \log (n/k) \log
n)$~\cite{IMS}. Very recently, a randomized protocol with a $O(k^2
\log n)$ bound on the number of bits transmitted was given in
\cite{CGK}. But the use of randomness makes these synchronization protocols
unsuitable for the application to deletion codes in the adversarial
model.

\subsection{Subsequent Work}

After posting of the preliminary version of this paper~\cite{self}, new results in the field of deletion codes have been found which either improve upon or complement our work.

Belazzougui \cite{B15} found a \emph{determiinstic} polynomial time algorithm for the document exchange problem for $k$ deletions with a message of length $O(k^2 + k\log^2(n))$. Most significantly, in this protocol, the number of deletions $k$ can be as large a $O(n^{1/3})$. The protocol is essentially a derandomization of the message length $O(k\log^2(n))$ protocol of \cite{IMS}. The hashes are derandomized using a \emph{deterministic sampling} procedure of Vishkin \cite{Vishkin91}. By tacking onto the original string this message (protecting it with the $(k+1)$-repetition code), achieves an efficient deterministic insertion/deletion code with message length of $O(k^3 + k^2\log^2(n))$.

Quite recently, Belazzougui and Zhang \cite{BZ16} found near-optimal efficient randomized constructions for a variety of problems related to the deletion channel when $k = O(n^{c})$ for some constant $c > 0$. For the document exchange problem, their procedure only needs $O(k(\log^{O(1)}(k) + \log(n))$ bits. There results also extend into the ``sketching'' problem, where Alice and Bob both generate hashes and send them to a third party which then computes all the necessary edits between the two strings, and the ``streaming problem,'' where the edit distance is to be computed with as little memory as possible under the condition Alice's string and then Bob's string are read in a stream. In particular, each of models these use $(k\log(n))^{O(1)}$ communication.

Since these latter constructions are randomized, our result is still the best-known deterministic deletion code in the regime $k$ is a constant.

\subsection{Organization}
In Section~\ref{sec:prelim}, we define the notation and describe some simple or well-known codes which will be used throughout the paper. Section~\ref{sec:hash-for-mixed} demonstrates how to efficiently encode and decode pattern rich strings against $k$-bit deletions using a hashing procedure. Section~\ref{sec:enc-to-mixed} describes how to efficiently encode any string as a pattern rich string. Section~\ref{sec:proof-of-thm} combines the results of the previous sections to prove Theorem~\ref{thm:main}. Section~\ref{sec:insertions} describes how to modify the code so that it works efficiently on the $k$-bit insertion and deletion channel. Section~\ref{sec:conclusion} suggests what would need to be done to improve redundancy past $O(k^2\log k\log n)$ using our methods.

\section{Preliminaries} \label{sec:prelim}

A subsequence of a string $x$ is any string obtained from $x$ by deleting one or more symbols. In contrast, a substring is a subsequence made of several consecutive symbols of $x$.

\begin{df}
Let $k$ be a positive integer. Let $\sigma_k : \{0,1\}^n \to 2^{\{0,1\}^{n-k}}$ be the function which maps a binary string $s$ of length $n$ to the set of all subsequences of $s$ of length $n - k$. That is, $\sigma_k(s)$ is the set of all possible outputs through the $k$-bit deletion channel. 
\end{df}

\begin{df}
Two $n$-bit strings $s_1$ and $s_2$ are \textit{$k$-confusable} if and only if $\sigma_k(s_1) \cap \sigma_k(s_2) \neq \emptyset$.
\end{df}

\def\hash{\mathsf{hash}}

We now state and develop some basic ingredients that our construction
builds upon. Specifically, we will see some simple constructions of
hash functions such that the knowledge $\hash(x)$ and an arbitrary
string $y \in \sigma_k(x)$ allows one to reconstruct $x$.  Our final
deletion codes will use these basic hash functions, which are either inefficient in terms of size or complexity, to build hashes that are efficient both in terms of size and computation time. These will then be used to build deletion codes, after protecting those hashes themselves with some redundancy to guard against
$k$ deletions, and then including them also as part of the codeword.

We start with an asymptotically optimal hash size which is inefficient to compute. For runtimes, we adopt the notation $O_k(f(n))$ to denote that the runtime may depend on a hidden function of $k$.

\begin{Lemma}\label{lem:brute-force} %
Fix an integer $k \ge 1$. There is a hash function $\hash_1 : \{0,1\}^n \to \{0,1\}^m$ for $m \le 2k\log n + O(1)$, computable in  $O_k(n^{2k}2^n)$ time, such that for all $x \in \{0,1\}^n$, given $\hash_1(x)$ and an arbitrary $y \in \sigma_k(x)$, the string $x$ can be recovered in $O_k(n^{2k}2^n)$ time.
\end{Lemma}
\begin{proof}
This result follows from an algorithmic modification of the methods of
\cite{Levenshtein66}. It is easy to see that for any $n$-bit string
$x$, $|\sigma_k(x)| \le n^k$. Additionally, for any $(n-k)$-bit string
$y$, the number of $n$-bit strings $s$ for which $y \in \sigma_k(s)$
is at most $2^k\binom{n}{k}\le 2n^k$. Thus, any $n$-bit string $x$ is
confusable with at most $2n^{2k}$ others strings. We view this as a graph on $\{0,1\}^n$, with an edge between two strings if they are confusable. We just showed that this graph has maximum degree at most $2n^{2k}$. Using the standard greedy procedure, one can $(2n^{2k}+1)$-color these strings in $O_k(n^{2k}2^n)$ time. 
 We can define $\hash_1(x)$ to be the color of $x$. 

Given such a hash and a $(n-k)$ bit received subsequence $y \in
\sigma_k(x)$, the receiver can in time $O_k(n^{2k}2^n)$ determine the color
of all strings $s$ for which $y \in \sigma_k(s)$. By design, exactly
one of these stings $s$ has the color $\hash_1(x)$; that is when $s = x$. So the receiver will be able to successfully decode $x$, as desired.
\end{proof}

We now modify the above result to obtain a larger hash that is however faster to compute (and also allows faster recovery from deletions).

\begin{Lemma}\label{lem:small-string}
Fix an integer $k \ge 1$.
There is a hash function $\hash_2 : \{0,1\}^n \to \{0,1\}^m$ for $m
\approx 2kn \log\log n/\log n$ computable in $O_k(n^2 (\log n)^{2k})$
time, such that for all $s \in \{0,1\}^n$, given $\hash_2(s)$ and an
arbitrary $y \in \sigma_k(s)$, the string $s$ can be recovered in
$O_k(n^2 (\log n)^{2k})$ time.
\end{Lemma}
\begin{proof}
We describe how to compute $\hash_2(s)$ for an input $s \in
\{0,1\}^n$.  Break up the string into consecutive substrings $s_1,
\hdots, s_{n'}$ of length $\lceil \log n \rceil$ except possibly for
$s_{n'}$ which is of length at most $\lceil \log n \rceil$. For each
of these strings, by Lemma~\ref{lem:brute-force} we can compute in $O_k(n(\log n)^{2k})$ time the string $\hash_1(s_i)$ of length $\sim 2k\log \log n$. Concatenating each of these hashes, we obtain a hash of
length $\sim 2kn\log \log n/\log n$ which takes $O_k(n^2(\log n)^{2k})$
time to compute. The decoder can recover the string $s'$ in
$O_k(n^2(\log n)^{2k})$ time by using the following procedure. For
each of $i \in \{1, \hdots, n'\}$ if $j_i$ and $j'_i$ are the starting
and ending positions of $s_i$ in $s$, then the substring between positions 
$j_i$ and $j'_i - k$ in $s'$ must be a subsequence of $s_i$.
Thus, applying the decoder described in Lemma $\ref{lem:brute-force}$, we can in
$O_k(n(\log n)^{2k})$ time recover $s_i$. Thus, we can recover $s$ in
$O_k(n^2(\log n)^{2k})$ time, as desired.
\end{proof}

We will also be using Reed-Solomon codes to correct $k$ symbol {\em errors}. For our purposes, it will be convenient to use a systematic version of Reed-Solomon codes, stated below. The claimed runtime follows from near-linear time implementations of unique decoding algorithms for Reed-Solomon codes, see for example \cite{Gao02}.


\begin{Lemma}\label{lem:RS} 
Let $k < n$ be positive integers, and $q$ be a power of two satisfying $n+2k \le q \le O(n)$. 
 Then
there exists a map $\RS : \F_q^n \to \F_q^{2k}$, computable in
$O_k(n(\log n)^4)$ time, such that the set $\{(x, \RS(x)) \mid x \in
\F_q^n\}$ is an error-correcting code that can correct $k$ errors in
$O_k(n(\log n)^4)$ time. In particular, given $\RS(x)$ and an arbitrary $z$ at
Hamming distance at most $k$ from $x$, one can compute
$x$ in $O_k(n (\log n)^4)$ time.
\end{Lemma}


\section{Deletion-correcting hash for \mixstr\ strings}
\label{sec:hash-for-mixed}
In this section, we will construct a short, efficiently computable hash that enables recovery of a string $x$ from $k$-deletions, when $x$ is typical in the sense that each short pattern occurs frequently in $x$ (we call such strings \emph{\mixstr}). \vnote{Should we give a more informative name, may be well-mixed or pattern-rich? I started a macro \mixstr\ but didn't change anything yet as perhaps this is not high priority.}

\subsection{Pattern-rich strings.}

We will use $n$ for the length of the (\mixstr) string to be hashed, and as always $k$ will be the number of deletions we target to correct. 
The following parameters will be used throughout:
\begin{align}
\label{eq:d-and-m}
d &=\floor{20000 k (\log k)^2\log n}\qquad \text{and}\\
m &= \lceil \log k+\log\log (k+1)+5 \rceil \ .
\end{align}
\noindent
It is easy to see that the choice of $m$ satisfies
\begin{equation}
\label{eq:m-and-k}
2^m > 2k (2m-1) \ . 
\end{equation}
Indeed, we have
\begin{align*}
  2^m &\ge 32k\log (k+1)\\
  &> 2k(15\log (k+1))\\
  &> 2k(2\log k + 2\log \log (k+1) + 11)\\
  &> 2k(2m-1) \ .
\end{align*}
\noindent
We now give the precise definition of mixed strings.
\begin{df}
Let $p$ and $s$ be binary strings of length $m$ and $n$, respectively, such that $m < n$. Define a $p$\textit{-split point} of $s$ be an index $i$ such that $p = s_is_{i+1}\hdots s_{i+m-1}$.
\end{df}

\begin{df}\label{def:mixed}  
We say that a string $s\in\{0,1\}^n$ is $k$-\textit{mixed} if for every $p\in\{0,1\}^m$, every substring of $s$ of length $d$ contains a $p$\textit{-split point}. Let $\mathcal M_n$ be the set of $k$-mixed strings of length $n$.
\end{df}

\subsection{Hashing of Mixed Strings.}
The following is our formal result on a short hash for recovering mixed strings from $k$ deletions.
\newcommand{\hashmixed}{\mathsf{hashMixed}}
\newcommand{\ph}{h_{\mathsf{pattern}}}
\newcommand{\gh}{g_{\mathsf{pattern}}}
\newcommand{\mixed}{\mathsf{mixed}}

\begin{theorem}\label{thm:hash}
Fix an integer $k \ge 2$. Then for all large enough $n$, there exists
$b = O(k^2 \log k \log n)$ and a hash function $H_\mixed : \mathcal
M_n \to \{0, 1\}^{b}$ and a deletion correction function $G_\mixed :
\{0,1\}^{n-k} \times \{0,1\}^b \to \{0,1\}^n \cup \{\mathsf{Fail}\}$,
  both computable in $O_k(n(\log n)^4)$ time, such that for any
  $k$-mixed $s\in\{0,1\}^n$, and any $s'\in \sigma_k(s)$, we have
  $G_\mixed(s',H_\mixed(s))=s$.
\end{theorem}

\begin{df}
\label{def:preserved-patterns}
If $s\in\{0,1\}^n$, $s'\in \sigma_k(s)$, and $p\in\{0,1\}^m$, we say that
$s'$ is $p$-\emph{preserving with respect to} $s$ if there are some
$1\le i_1\le\ldots\le i_k\le n$ such that $s'$ is obtained from $s$ by
deleting $s_{i_1},\ldots,s_{i_k}$ and:
\begin{enumerate}
\itemsep=0ex
\item no substring of $s$ equal to $p$ contains any of the bits at positions $i_j$
\item $s$ and $s'$ have an equal number of instances of substrings equal to $p$
\end{enumerate}
\end{df}
Intuitively, $s'$ is $p$-preserving with respect to $s$ if we can obtain $s'$ from $s$ by deleting $k$ bits without destroying or creating any instances of the pattern $p$.

We first prove the following lemma. 
\begin{Lemma}\label{thm:hashforonep}
Fix an integer $k \ge 2$. Then for sufficiently large $n$, there exists a hash
function 

$\quad \ph: \mathcal M_n \times \{0,1\}^m \to \{0,1\}^{2k(\lceil\log
  n\rceil+1)}$ and deletion correction function 
  
  $\quad \gh: \{0,1\}^{n-k} 
\times \{0,1\}^{2k(\lceil\log n\rceil+1)} \times \{0,1\}^m\to \{0,1\}^n \cup
\{\Fail\}$, 

both computable in $O_k(n(\log n)^4)$ time, such that for
every pattern $p\in\{0,1\}^m$, every $k$-mixed $s\in\{0,1\}^n$, and an arbitrary
$s'\in \sigma_k(s)$ that is $p$-preserving with respect to $s$, one has 
\[ \gh(s',\ph(s,p),p)=s \ . \]
\end{Lemma}
\begin{proof}

We first define the hash function $\ph$ as follows.  
Assume we are
given a mixed string $r\in\mathcal M_n$ and a pattern $p\in \{0,
1\}^m$. Let $a_1,\ldots,a_u$ be the $p$-split points of $r$.
Then we let strings $w_0,\ldots,w_u$ be defined by $w_0=r_0\cdots
r_{a_1-1}$, $w_u=r_{a_u}\cdots r_{n-1}$, and for $1\le i\le u-1$,
$w_j=r_{a_j}\cdots r_{a_{j+1}-1}$. Thus $\{w_j\}$ are the strings that
$r$ is broken into by splitting it at the split points. By the
definition of a mixed string, each $w_j$ has length at most $d$ (as
defined in \eqref{eq:d-and-m}).

We let $\ell_j$ be the length of $w_j$, let $v_j$ be $w_j$ padded to
length $d$ by leading 0's, and let $y_j = \hash_2(v_j)$ as defined in Lemma~\ref{lem:small-string}, with the binary representation of the length of $\ell_j$
appended. We can compute $y_j$ in time $O_k((\log n)^2(\log \log
n)^{2k})$ and $y_j$ has length $v$ satisfying
\begin{align}
\label{eq:len-of-hash}
v&=O\left( 2k(k\log k)^2\log n \frac{\log\log\log n}{\log \log n}+\ceil{\log d}
\right)\\ \nonumber &< \log n \ 
\end{align}
for large enough $n$.

Let $x_j$ be the number whose binary representation is $y_j$. Then based on the length of $y_j$, we have that $x_j<n$. 
Let $q$ be the smallest power of $2$ that is at least $n+2k$. 
We then apply lemma~\ref{lem:RS} to $x=(x_1,\ldots,x_n) \in \F_q^n$ (with all those that are not defined being assigned value $0$) to obtain $(y_1,\ldots,y_{2k}) = \RS(x) \in \F_q^{2k}$. For $1\le j \le 2k$, let $S_j$ be the binary representation of $y_k$, padded with leading 0's so that its length is $\lceil \log n \rceil+1$.

Finally, we define the hash value 
\[ \ph(s,p) = S_1\cdots S_{2k} \ . \]
Clearly, the length of $\ph(s,p)$ equals $2k (\lceil \log n \rceil + 1)$.

To compute $\gh(s',\tilde{h},p)$, where $s'$ is a subsequence of $s$ that is $p$-pattern preserving with respect to $s$, we split $\tilde{h}$ into $2k$
equal-length blocks, calling them $S_1,\ldots, S_{2k}$. We compute
$(x'_1,\ldots,x'_n)$ from $s'$ in the same way that we computed
$(x_1,\ldots,x_n)$ from $s$ when defining $\ph(s,p)$. Now, assuming
$\tilde{h}=\ph(s,p)$, there are at most $k$ values of $j$ such that
$x'_j\ne x_j$, since there are at most $k$ deletions.
We can use Lemma~\ref{lem:RS} and $S_1,\ldots, S_{2k}$ to
correct these $k$ errors. From a corrected value of $x_j$, we can
obtain the value of $w'_j$ and $\ell_j$. Since $\ell_j$ is the length
of $w_j$, we can use it to remove the proper number of leading zeroes
from $w'_j$ and obtain $w_j$. Thus we can restore the original $s$ in
$O_k(n(\log n)^4 + n(\log n)^2(\log \log n)^{2k})$ time. Since $(\log \log n)^{2k} \le O_k(1) + O(\log n)$,\footnote{For instance, $(\log \log n)^{2k} \le O(2^{2k^2} + \log n)$, because either $\log \log n \le 2^k$ in which case $(\log \log n)^{2k} \le 2^{2k^2}$, or $(\log \log n)^{2k} \le (\log \log n)^{2 \log \log \log n} \le O(\log n)$.}%
~the overall decoding time is $O_k(n (\log n)^4)$.
\end{proof}

\noindent
With the above lemma in place, we are now ready to prove the main theorem of this section.

\begin{proof}{(of Theorem~\ref{thm:hash})}
Given a mixed string $s \in \mathcal{M}_n$, the hash $H_\mixed(s)$ is computed
by computing $\ph(s,p)$ from Lemma~\ref{thm:hashforonep} for each
pattern $p\in\{0,1\}^m$ and concatenating those hashes in order of
increasing $p$. For the decoding, to compute $G_\mixed(s',
H_\mixed(s))$ for a $s' \in \sigma_k(s)$, we run $\gh$ from
Lemma~\ref{thm:hashforonep} on each of the $2^m$ subhashes
corresponding to each $p \in \{0,1\}^m$, and then take the
majority (we can perform the majority bitwise so that it runs in $O_k(n)$ time).
When deleting a bit from $s$, at most $(2m-1)$ patterns $p$
are affected (since at most $m$ are deleted and at most $m-1$ are
created). Thus $k$ deletions will affect at most $k(2m-1)$ patterns of
length $m$. Since $m$ was chosen such that $2^m> 2k(2m-1)$, we have
that $s'$ is $p$-preserving with respect to $s$ for a majority of
patterns $p$. Therefore, we will have $\gh(s',\ph(s,p),p)=s$ for a
majority of patterns $p$, and thus $G_\mixed$ reconstructs the string
$s$ correctly.
\end{proof}

\section{Encoding into Mixed Strings} 
\label{sec:enc-to-mixed}
The previous section describes how to protect mixed strings against
deletions.  We now turn to the question
of encoding an arbitrary input string $s \in \{0,1\}^n$ into a mixed string in $\mathcal{M}_n$. 

\begin{df}
Let $\mu : \{0, 1\}^n \times \{0, 1\}^L \to \{0, 1\}^n$ be the function which takes a string $s$ of length $n$ and a string $t$, called the \textit{template}, of length $L$, and outputs the bit-wise XOR of $s$ with $t$ concatenated $\lceil n / L \rceil$ times and truncated to a string of length $n$.
\end{df}

\noindent We will apply the above function with the parameter choice
\begin{equation}
\label{eq:def-of-L}
 L=\ceil{m2^m(\log(n2^m)+1)}\le\ceil{10000k(\log k)^2\log n}-1 \ .
\end{equation}
\noindent 
Equation \ref{eq:def-of-L} follows since for $k \ge 2$
\begin{align*}
  \lceil m 2^m(\log (n2^m) + 1)\rceil &\le (\log k + \log \log (k+1) + 6)\\
  & \quad(64 k\log (k+1))(\log n + m + 1)\\
  &\le (8\log k)(128k \log k)(2\log n)\\
  &\le \lceil 10000k(\log k)^2(\log n)\rceil - 1
\end{align*}

Notice that for all $s \in \{0, 1\}^n$ and $t \in \{0, 1\}^L$, $\mu(\mu(s, t), t) = s$. Notice also that $\mu$ is computable in $O(n)$ time.
It is not hard to see that for any $s \in \{0,1\}^n$, the string
$\mu(s,t)$ for a random template $t \in \{0,1\}^L$ will be $k$-mixed
with high probability. We now show how to find one such template $t$ that is suitable for $s$, 
deterministically in near-linear time.

\begin{Lemma}\label{lem:mixed-string}
There exists a function $T : \{0, 1\}^n \to \{0, 1\}^L$ such that for all $s \in \{0, 1\}^n$, $\mu(s, T(s)) \in \mathcal M_n$. Also, $T$ is computable in $O(k^3(\log k)^3 \ n\log n) = O_k(n\log n)$ time.
\end{Lemma}
\begin{proof}
For a given string $s$ and template $t$, let $r =
\mu(s, t)$. We say that a pair $(i,p)\in \{0,1,\dots,\floor{n/L}-1\} \times \{0,1\}^m$ is an \emph{obstruction} if the substring of $r_{iL+1}\cdots r_{iL+L}$
does not include the pattern $p$ as a substring. We will construct $t=T(s)$ so that $r$ has no obstructions, and then $r$ is mixed.

We will choose $T(s)$ algorithmically. For $0\le j\le \floor{L/m}-1$, at the beginning of step $j$ we will have $b_j$ potential obstructions. Clearly, $b_0=\floor{n/L}2^m$. At step $j$, we will specify the values of $t_{jm+1}\cdots t_{jm+m}$. If we chose these bits randomly, then $\mathbb{E}[b_{j+1}]=(1-2^{-m})b_j$. Thus we can check all of the $2^m$ possibilities and find some way to specify the bits so that $b_{j+1}\le (1-2^{-m})b_j$. This will then give us that
\[
b_{\floor{L/m}}\le (1-2^{-m})^{\floor{L/m}}(\floor{n/L}2^m)\le e^{-\floor{L/m}2^{-m}}n2^m
\]
which is less than $1$ 
by our choice of $L$ in \eqref{eq:def-of-L}.

Thus we can find a template $T(s)$ with no obstructions in $\floor{L/m}$ steps. The definition of an obstruction tells us that every substring of $r = \mu(s, T(s))$ of length $2L\le\floor{20000k(\log k)^2\log n} = d$ contains every pattern of length $m$, so the string $r \in \mathcal M_n$.

Let us estimate the time complexity of finding $T(s)$. Fix a step $j$, $0 \le j < \floor{L/m}$. Going over all possibilities of  $t_{jm+1}\cdots t_{jm+m}$ takes $2^m$ time, and for each estimating the reduction in number of potential obstructions takes $O(n 2^m)$ time.  The total runtime is thus $O(n 4^m L/m) = O(n 8^m \log (n 2^m)) = O(k^3 (\log k)^3 n \log n)$.
\end{proof}

\section{The Encoding/Decoding Scheme: Proof of Theorem \ref{thm:main}}\label{sec:proof-of-thm}

Combining the results from Sections \ref{sec:hash-for-mixed} and
\ref{sec:enc-to-mixed}, we can now construct the encoding/decoding functions $\enc$ and
$\dec$ which satisfy Theorem \ref{thm:main}. But first, as a warm-up,
we consider a simple way to obtain such maps for codewords of slightly larger length $N \le n + O(k^3\log k\log n)$ (i.e., the redundancy has a cubic rather than quadratic dependence on $k$). Let $s$ be the message string we seek to
encode. First compute $t = T(s)$ and $r = \mu(s, t)$ as per Lemma~\ref{lem:mixed-string}. Then, define the encoding of $s$ as
\[ \enc(s)= \langle r, \
\rep_{k+1}(t), \ \rep_{k+1}(H_\mixed(r) \ \rangle  \ , \] where $H_\mixed(\cdot)$ is the deletion-correcting hash function for mixed strings from Theorem~\ref{thm:hash}, and  $\rep_{k+1}$ is the
$(k+1)$-repetition code which repeats each bit $(k+1)$ times. Since $H_\mixed(r)$ is the most intensive computation,
$\enc(s)$ can be computed in $O_k(n(\log n)^4)$ time and its length
is
\begin{align*}
  n &+ (k+1)L + (k+1) |H_\mixed(r)|\\
  &\le n + (k+1)O(k\log^2 k\log n) + (k+1)O(k^2\log k\log n)\\
    &\le  n + O(k^3\log k\log n).
\end{align*}
We now describe the efficient decoding function $\dec$. Suppose we
receive a subsequence $s' \in \sigma_k(\enc(s))$, First, we can easily
decode $t$ and $H_\mixed(r)$, since we know the lengths of $t$ and
$H_\mixed(r)$ beforehand. Also the first $n-k$ symbols of $s'$ yield a
subsequence $r'$ of $r$ of length $n-k$. Then, as shown in
Theorem~\ref{thm:hash}, $r = G_\mixed(r', H_\mixed(r))$ and this can
be computed in $O_k(n (\log n)^4)$ time. Finally, we can compute $s =
\mu(r, t)$, so we have successfully decoded the message $s$ from $s'$,
as desired.

\smallskip
Now, we demonstrate how to obtain the improved encoding length of $n + O(k^2\log k\log n)$. The idea is to use Lemma \ref{lem:small-string} to protect $t$ and $H_\mixed(r)$ with less redundancy than the naive $(k+1)$-fold repetition code.

\begin{proof}{(of Theorem~\ref{thm:main})}
Consider the slightly modified encoding
\begin{align}
\label{eq:final-encoding}
\enc(s) = \langle r, &\ t, \ \rep_{k+1}(\hash_2(t)),\ H_\mixed(r),\\
\nonumber &\ \rep_{k+1}(\hash_2(H_\mixed(r))) \ \rangle \ . 
\end{align}
The resulting codeword can be verified to have length $O(k^2(\log
k)(\log n))$ for large enough $n$; the point is that $\hash_2()$
applied to $t$ and $H_\mixed(r)$, which are $O_k(\log n)$ long
strings, will result in strings of length $o_k(\log n)$, so we can
afford to encode them by the redundancy $(k+1)$ repetition code,
without affecting the dominant $O_k(\log n)$ term in the overall
redundancy.

Since we know beforehand, the starting and ending positions of each of
the five segments in the codeword \eqref{eq:final-encoding}, we can in
$O(n)$ time recover subsequences of $r$, $t$,
$\rep_{k+1}(\hash_2(t))$, $H_\mixed(r)$, and
$\rep_{k+1}(\hash_2(H_\mixed(r))$ with at most $k$ deletions in
each. By decoding the repetition codes, we can recover $\hash_2(t)$
and $\hash_2(H_\mixed(r))$ in $O(n)$ time. Then, using the algorithm
described in Lemma \ref{lem:small-string}, we can recover $t$ and
$H_\mixed(r)$ in $O_k(n'^2(\log n')^{2k})$ time where $n' = \max (L,
|H_\mixed(r)|) = O(k^2\log k\log n)$. 
Once $t$ and $H_\mixed(r)$ are recovered, we can proceed as in the 
previous argument and decode $s$ in $O_k(n(\log n)^4)$ time, as
desired. 
\end{proof}


\section{Efficient Algorithm for Correcting Insertions and Deletions}\label{sec:insertions}


By a theorem of Levenshtein \cite{Levenshtein66}, we have that our code works not only on the $k$-bit deletion channel but also on the $k$-bit insertion and deletion channel. The caveat though with this theorem is that the decoding algorithm may not be as efficient. In this section, we demonstrate a high-level overview of a proof that, with some slight modifications, the code we constructed for $k$ deletions can be efficiently decoded on the $k$-bit insertion and deletion channel. Although the redundancy will be slightly worse, its asymptotic behavior will remain the same.

To show that our code works, we argue that suitable modifications of each of our lemmas allow the result to go through.

\begin{itemize}
\vspace{-1ex}
\item Lemma \ref{lem:brute-force} works for the $k$-bit insertion and deletion channel by Levenshtein's result \cite{Levenshtein66}. Since encoding/decoding were done by brute force the efficiency will not change by much.

\item To modify Lemma \ref{lem:small-string} we show that the code which corrects $3k$ deletions can also correct $k$ insertions and $k$ deletions nearly as efficiently. If the codeword transmitted is $s_1, \hdots , s_{n'}$ where each $s_i$ is of length at most $\lceil \log n\rceil$ then in the received word, if $s_i$ was supposed to be in positions $i_a$ to $i_b$, then positions $i_a + k$ to $i_b - k$ must contain bits from $s_i$ except possibly for $k$ spurious insertions. Using Lemma \ref{lem:brute-force} modified for insertions, we can restore $s_i$ using brute force, and thus we can restore the original string with about the same runtime as before.

\item Lemma \ref{lem:RS} and Lemma \ref{thm:hashforonep} do not change because the underlying error-correcting code does not depend on deletions or insertions.

\item Theorem \ref{thm:hash} extends because the number of patterns $p$ which are preserved (in the sense of Definition~\ref{def:preserved-patterns})  with respect to a specific pattern of $k$ insertions and deletions is roughly the same as with $k$ deletions. And, for such a preserved pattern, the associated hash function $\ph$ allows for correction of arbitrary bounded number of errors in strings between the $p$-split points, and it doesn't matter if those errrors are created
by insertions or deletions.


\item Lemma \ref{lem:mixed-string} does not change.

\item Theorem \ref{thm:main} needs some modifications. The encoding has the same general structure except we use the hash functions of the modified lemmas and we use a $(3k+1)$-fold repetition code instead of a $(k+1)$-fold repetition code. That is, our encoding is 
\begin{align*}\phi(s) = \langle r,&~ t,~ \rep_{3k+1}(\hash_2(t)),~ H_\mixed(r),\\&~ \rep_{3k+1}(\hash_2(H_\mixed(r))) \rangle \ .\end{align*}
In the received codeword we can identity each section with up to $k$ bits missing on each side and $k$ spurious insertions inside. In linear time we can correct the $(3k+1)$-repetition code by taking the majority vote on each block of length $3k+1$. Thus, we will have $\hash_2(t)$ and $\hash_2(H_\mixed(r))$ from which we can obtain $t$, $H_\mixed(r)$, and finally $r$ and $s$ in polynomial (in fact, near-linear) time as in the deletions-only case.

\end{itemize}
Thus, we have exhibited an efficient code encoding $n$ bits with $O(k^2 \log k \log n)$ redundancy and which can be corrected in near-linear time against any combination of insertions and deletions totaling $k$ in number.

\section{Concluding remarks}\label{sec:conclusion}

In this paper, we exhibit a first-order asymptotically optimal
efficient code for the $k$-bit deletion channel. Note that to improve
the code length past $n + O(k^2 \log k \log n)$, we would need to modify our hash
function $H_\mixed(r)$ so that either it would use shorter hashes for each
particular pattern $p$ or it would require using fewer patterns
$p$. The former would require distributing the hash information
between different patterns, which may not be possible since the
patterns do not synchronize with each other. An approach through the
latter route seems unlikely to improve past $n + O(k^2 \log n)$ since
an adversary is able to ``ruin" $k$ essentially independent patterns
because the string being transmitted is $k$-mixed.

Another interesting challenge is to give a deterministic one-way protocol with $\mathrm{poly}(k \log n)$ communication for synchronizing a string $x \in \{0,1\}^n$ with a subsequence $y \in \sigma_k(x)$ (the model discussed in Section~\ref{sec:sync}). The crux of our approach is such a protocol when the string $x$ is mixed, but the problem remains open when $x$ can be an arbitrary $n$-bit string.

Another intriguing question is whether there is an extension of the Varshamov-Tenengolts (VT) code for the multiple deletion case, possibly by using higher degree coefficients in the check condition(s) (for example, perhaps one can pick the code based on $\sum_{i=1}^n i^a x_i$ for $a=0,1,2,\dots,d$ for some small constant $d$). Note that this would also resolve the above question about a short and efficient deterministic hash for the synchronization problem. However, for the case of two deletions there are counterexamples for $d \le 4$, and it might be the case that no such bounded-degree polynomial hash works even for two deletions.

\section*{Acknowledgments}

The second author thanks Michael Saks for valuable discussions about the problem of recovering from  multiple deletions and in particular about the possibility of finding a VT-like code for correcting two deletions. We are grateful to Kuan Cheng and Xin Li for pointing out an error in the appendix of the published versions of our paper~\cite{self,self2}, namely that the claim about the rate limitation of linear $k$-bit deletion codes is false.

\bibliographystyle{alpha}
\bibliography{cite2}

\end{document}